# Superionic Helium in $Mg_2SiO_4He_2$ at Ultrahigh Pressure: Implications for Delayed Cooling of Hot Jupiters

Wenxuan Chen,[1,2] Shumin Guo,[1] Chengda Li,[1] Yingze Wang,[1,2] Jialiang Cai,[1,2] Guiyan Dong,[1] Yangfan Gao,[2] Jing Guo,[2,3,*] Simon A. T. Redfern,[4,5†] and Defang Duan[1,‡]

[1]*Key Laboratory of Material Simulation Methods & Software of Ministry of Education and State Key Laboratory of High Pressure and Superhard Materials, College of Physics, Jilin University, Changchun 130012, China*

[2]*Institute of Physics, Chinese Academy of Sciences, Beijing 100190, China*

[3]*University of Chinese Academy of Sciences, Beijing 100049, China*

[4]*Asian School of the Environment, Nanyang Technological University, Singapore 639798, Singapore*

[5]*School of Materials Science and Engineering, Nanyang Technological University, Singapore 639798, Singapore*

**ABSTRACT**. The radius inflation of hot Jupiters remains a long-standing puzzle. Existing explanations invoke physical heating or delayed cooling through enhanced atmospheric opacity or inhibited convection, yet the possible role of reactions between volatiles and silicates under extreme conditions has received little attention. Here we predict a previously unknown compound, $Mg_2SiO_4He_2$, that becomes thermodynamically stable above 577 GPa, exhibits superionic helium diffusion at high temperature, and has thermal conductivity approximately one-quarter that of $Mg_2SiO_4$. Its formation at hot Jupiter core-mantle boundaries sequesters helium into the silicate, creating a compositionally stratified, insulating thermal boundary layer that delays planetary cooling, a chemical mechanism complementing existing models of radius inflation. Our results provide new constraints on the chemical environment and thermal evolution of giant exoplanetary interiors.

## Introduction

Hot Jupiters are a class of giant planets with extremely short orbital periods and equilibrium temperatures exceeding 1000 K[1]. Their structure primarily consists of an outer atmospheric layer, a molecular hydrogen layer, a metallic hydrogen layer, and an inner core. In sufficiently cool giant planets, helium from the envelope can be transported into the metallic hydrogen layer as helium rain. Since the discovery of the first transiting hot Jupiter, HD 209458 b, its widespread "radius inflation", where the actual radius is 20% to 80% larger than predicted by standard evolutionary models, has remained one of the central mysteries in planetary science[2, 3]. Observational data indicate that this degree of expansion is strongly correlated with the stellar radiation flux received by the planet, suggesting that the energy source is closely tied to the host star[4-6].

A variety of physical processes have been invoked in an attempt to explain the observed radius inflation of hot Jupiters. The mainstream explanations fall into two categories: a kinetic heating mechanism and a thermal barrier mechanism. The former includes tidal heating and ohmic dissipation[7, 8], while the latter emphasizes the delayed cooling effect caused by increased atmospheric opacity[9]. However, existing models face a critical challenge: how does energy effectively accumulate in the deep adiabatic layer? Even if the upper atmosphere is strongly heated, energy deposited above the radiative-convective boundary is efficiently re-radiated to space and contributes little to the planet's overall radius; heating must reach sufficient depth to raise the interior entropy and slow contraction[10]. Recent studies suggest that the dominant heating mechanism may vary with temperature, and existing mechanisms are unlikely to account for all observations[11, 12].

Meanwhile, advances in high-pressure physics have provided a completely new perspective[13-27]. Recent studies have revealed that under extreme pressures, rare gases previously considered chemically inert (such as He and Ar) can form stable compounds with metals, ices, and silica, including FeHe, He-$SiO_2$, He-$NH_3$, and He-$H_2O$[28-36]. These findings suggest that hitherto unrecognized rare-gas reservoirs exhibit unique transport properties at high temperatures. On the other hand, significant progress has been made in the study of hydrated/hydrogen silicates[19, 37-40]. These studies highlight the complex interactions between volatiles and silicates under extreme conditions and demonstrate their potential impact on planetary thermal evolution and internal structures.

In this study, we predict a stable helium-silicate compound based on forsteritic olivine stoichiometry, $Mg_2SiO_4He_2$, and propose an entirely new mechanism for the radius inflation of hot Jupiters. The chemical incorporation of helium into the silicate at depth creates a compositional boundary that may inhibit large-scale convection, providing a previously unrecognized chemical explanation for delayed cooling. Our findings not only fill a critical gap in the silicate-helium system but also offer a new perspective on the factors governing the thermal evolution and observed radius anomalies of hot Jupiters.

# Results

Considering that the masses of hot Jupiters range from approximately 0.3 to 10 $M_J$, the temperature and pressure conditions vary significantly in hot Jupiters with different masses. Therefore, the models of HD 209458 b, with a mass of 0.69 $M_J$, were selected for subsequent studies[41, 42], with deep-interior conditions comparable to those of multi-layer Jupiter models[43]. We performed structural searches of the $Mg_2SiO_4$-He binary system at 500, 800 and 1000 GPa, and constructed the thermodynamic convex hull, as shown in Fig. 1a. At 500 GPa, two stoichiometric compounds, $Mg_2SiO_4He_2$ and $Mg_2SiO_4He$, lie on the convex hull, indicating they are thermodynamic stable. When the pressure increases to 800 GPa, $Mg_2SiO_4He$ deviates from the convex hull and becomes metastable, whereas $Mg_2SiO_4He_2$ remains stable. This stability of $Mg_2SiO_4He_2$ persists up to 1000 GPa, where all other stoichiometric compounds have deviated from the convex hull. Consequently, $Mg_2SiO_4He_2$ is the only thermodynamically stable compound in the pressure range of 800-1000 GPa. Figure 1b illustrates the pressure dependence of the formation enthalpy of $Mg_2SiO_4He_2$ relative to different decomposition pathways. By comparing the relative enthalpies of the predicted compounds with those of elemental helium, periclase, silicon dioxide and previously reported Mg-bearing silicates (e.g., $MgSiO_3$, $Mg_2SiO_4$), we confirm the thermodynamic stability of $Mg_2SiO_4He_2$ across the pressure range of 577-1000 GPa. Above 577 GPa, its enthalpy is significantly lower than that of all competing phase mixtures. This suggests that in hot Jupiters, when helium comes into contact with the rock core, it can react with silicates $Mg_2SiO_4$ in the core via $Mg_2SiO_4 + 2He \rightarrow Mg_2SiO_4He_2$.

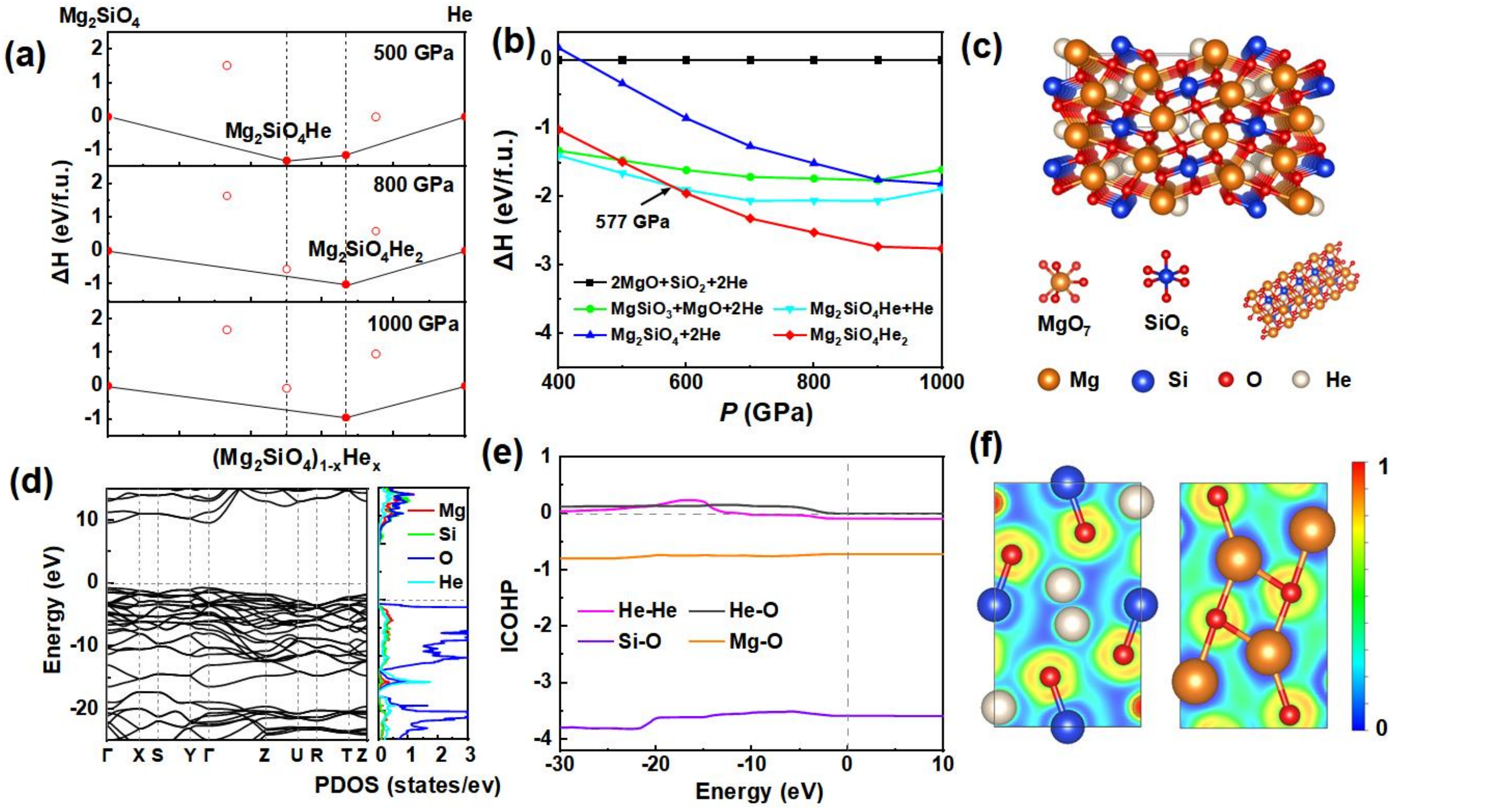

**Fig.1 The thermodynamic stability, crystal structure, and electronic structure of $Mg_2SiO_4He_2$. a** Convex hull diagram at 500, 800 and 1000 GPa. **b** Pressure dependence of the formation enthalpies. **c** Crystal structure of $Mg_2SiO_4He_2$. **d** The electronic band structure, PDOS and **e** ICOHP at 800 GPa. **f** 2D-ELF on the (001) plane.

$Mg_2SiO_4He_2$ crystallizes in the *Pbam* space group. As shown in Fig. 1c, its lattice consists of a framework of $MgO_7$ and $SiO_6$ coordination polyhedra, with the Mg-Si-O sublattice forming quasi-one-dimensional chains and the He atoms occupying the intervening voids. The distance between two adjacent He atoms is 1.08 Å, slightly shorter than the atomic distance in elemental helium (1.29 Å) at 800 GPa. Electronic properties reveal that $Mg_2SiO_4He_2$ is a wide-bandgap insulator with a bandgap of approximately 10 eV (Fig. 1d), comparable to that of $Mg_2SiO_4$. Weak hybridization between He 1s orbitals and O 2p orbitals occurs at -3 eV relative to the Fermi level, indicating minimal polarizing interactions rather than typical covalent or ionic bonds.

Integrated crystal orbital Hamilton population (ICOHP) and electron localization function (ELF) analyses provide further insight into the bonding characteristics. For Si-O bonds, the ICOHP reaches approximately -4 (Fig. 1e), indicating strong bonding, while the ELF is about 0.8 (Fig. 1f), consistent with pronounced covalent character. In contrast, the Mg-O interactions are predominantly ionic and weaker than the Si-O bonds, as reflected by their lower ELF (~0.5) and smaller-magnitude ICOHP (-0.72) values. The He atoms exhibit little direct bonding with neighboring atoms. The ICOHP values for He-He and He-O pairs are nearly zero, and the corresponding ELF values remain below 0.2, indicating that even under ultra-high pressure, the interactions between He and surrounding atoms exhibit closed-shell characteristics. Phonon calculations at 500, 800 and 1000 GPa show the absence of any imaginary frequencies, confirming the dynamical stability of $Mg_2SiO_4He_2$ over this pressure range [see Fig. S1 in Supplemental Information (SI)]. Similar to $Mg_2SiO_4He_2$, the crystal structure of $Mg_2SiO_4He$ (Fig. S3) features a Mg-Si-O framework that forms one-dimensional channels for He atoms. However, its thermodynamic stability field differs from the pressure range expected at the core-mantle boundaries of hot Jupiters, and we therefore do not consider this phase further.

To investigate the dynamical behavior of $Mg_2SiO_4He_2$, we performed machine-learning molecular dynamics (MLMD) simulations over a pressure range of 500-1000 GPa and a temperature range of 4000-11000 K. We characterized temperature- and pressure-induced phase transitions by analyzing the mean square displacement (MSD) and the diffusion coefficients of all atomic species. At 800 GPa, the species-dependent diffusion behavior revealed four distinct phase regimes.

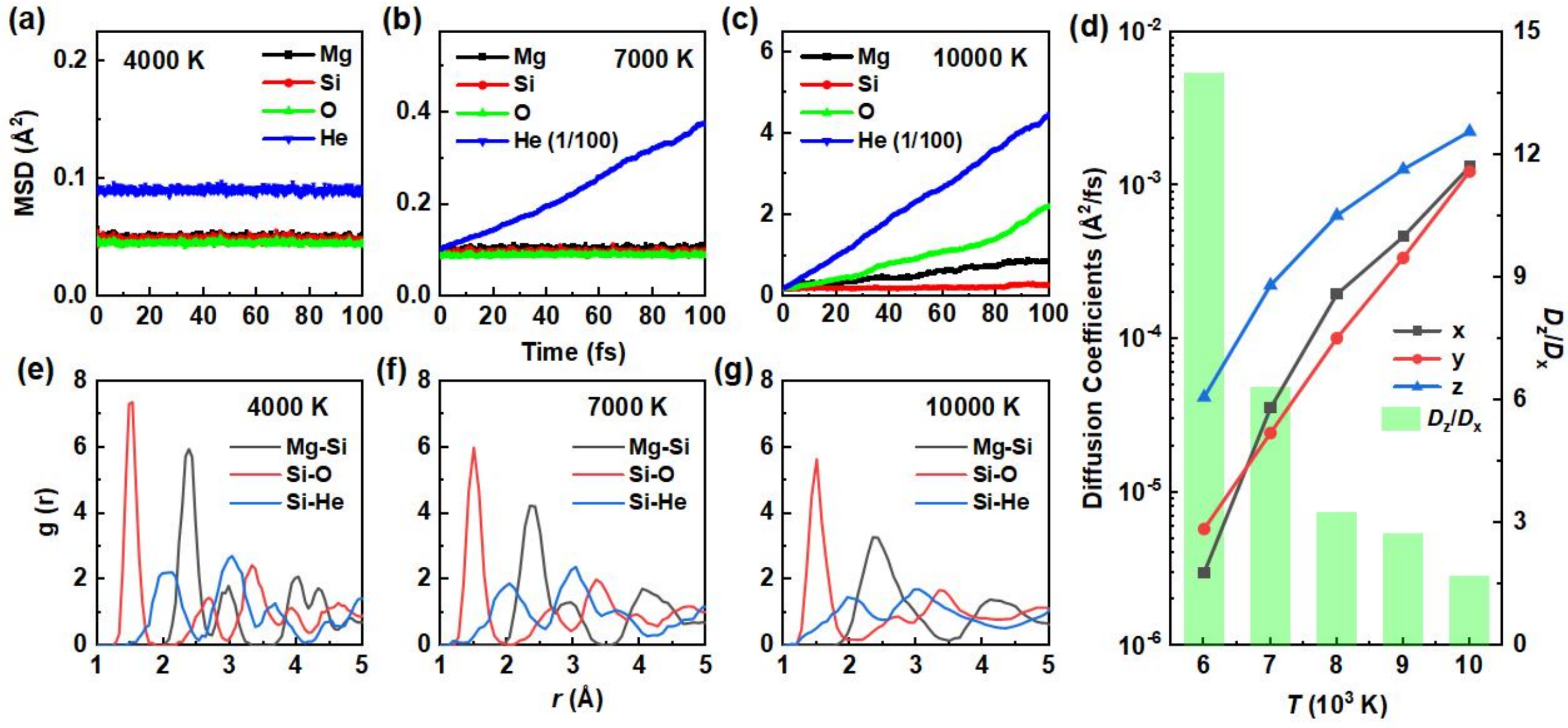

**Fig.2** The MSD at 800 GPa and at **a** 4000, **b** 7000 and **c** 10000 K. Due to the intense diffusion, Fig. 2b and 2c display only 1/100 of the MSD values of He. **d** Temperature-dependent diffusion coefficients of He in different

directions at 800 GPa and the anisotropic parameter $D_z/D_x$ in $Mg_2SiO_4He_2$. The RDFs at 800 GPa and at **e** 4000, **f** 7000 and **g** 10000 K.

For T < 5600 K (phase I), $Mg_2SiO_4He_2$ remains in the solid state. Within this range, the MSD of Mg, Si, O, and He plateau over time (Fig. 2a), with atoms vibrating near their equilibrium positions, indicating a crystalline solid. As shown in Fig. 2e, the radial distribution functions (RDF) exhibit sharp characteristic peaks, providing further evidence of long-range crystalline order.

For 5600 K ≤ T < 9600 K, $Mg_2SiO_4He_2$ enters phase II, corresponding to the superionic state of He. As illustrated in Fig. 2b, the MSD of He atoms increases linearly with time at 7000 K, indicating long-range diffusive motion. At 7000 K and 800 GPa, the diffusion coefficient $D_{He}$ reaches approximately $2.82\times10^{-4}$ Å$^2$/fs, while the MSDs of Mg, Si, and O plateau, indicating that these species remain localized near their equilibrium lattice sites. These results identify phase II as a He-superionic state. Notably, the Mg-Si-O framework provides a clear diffusion pathway for He. As shown in Fig. 2d, He diffusion exhibits strong anisotropy, with the diffusion coefficient along the c-direction being approximately one order of magnitude larger than those along a- and b-directions. This behavior reflects quasi-one-dimensional He motion constrained by the lattice framework. This anisotropy is strongest near the onset of superionicity, and as temperature rises further, diffusion in the a- and b-directions gradually increases while the anisotropy diminishes (e.g., at 6000 K, $D_z/D_x \approx 12$; at 7000 K, it decreases to ≈ 6). At this stage, enhanced thermal vibrations broaden the RDF peaks for Mg-Si and Si-O pairs, though the framework retains short-range order. The first- and second-neighbor peaks for the Si-He RDF persist, whereas the longer-range Si-He correlations disappear (Fig. 2f).

Above 9600 K, $Mg_2SiO_4He_2$ enters phase III, a second superionic regime. At this stage, the Mg and O MSDs increase linearly with time (Fig. 2c), indicating the onset of long-range diffusion. Figure 2g shows that the first- and second- neighbor RDF peaks broaden significantly, and that the two Mg-Si peaks beyond 4 Å merge. The first-neighbor peak of Si-O remains sharp, whereas the two peaks between 2 and 4 Å become markedly broad. Strong He diffusion is accompanied by pronounced broadening of the first-neighbor Si-He RDF peak. Upon further heating above 10300 K, $Mg_2SiO_4He_2$ fully melts and transitions into liquid phase IV.

We extended the pressure range to 500-1000 GPa, as shown in Fig. 3a. Each colored symbol represents an independent MLMD simulation result to exclude memory effects. Based on MSD and crystal structure, we identified four final states consistent with those at 800 GPa: solid phase I (red square), superionic phase II (green diamond), superionic phase III (cyan triangle), and liquid phase IV (orange circle). These classifications delineate the corresponding regions of the phase diagram.

Across the entire pressure range, the transition to phase II occurs at temperatures more than 5000 K below the $MgSiO_3$ melting curve[44]. Transitions to phases III and IV likewise occur well below this curve. The resulting phase diagram suggests that, under the pressure-temperature conditions expected at hot-Jupiter core-mantle boundaries, $Mg_2SiO_4He_2$ exists in superionic phase III. Because superionic states often exhibit reduced thermal conductivity, these phases raise the possibility that $Mg_2SiO_4He_2$ conducts heat less efficiently than $Mg_2SiO_4$ under comparable conditions. To test this possibility, we calculated the temperature-dependent thermal conductivities of both compounds (see methods in SI) under 800 GPa, as shown in Fig. 3b. At 4000 K, $Mg_2SiO_4He_2$ and $Mg_2SiO_4$ are both solid and have comparable thermal conductivities of 23.9 and 25.8 W/mK, respectively. At 5000 K, the thermal conductivities of both compounds decrease slightly. Upon further heating, $Mg_2SiO_4He_2$ first transforms into the superionic state, with He atoms rapidly diffusing and causing an abrupt decrease in thermal conductivity, which plummets to 6.1 W/mK at 6000 K. At the same temperature and pressure, this value is only one-quarter of that of $Mg_2SiO_4$. When the temperature reaches 10000 K, the thermal conductivity of $Mg_2SiO_4He_2$ undergoes continuous decline to 3 W/mK. In contrast, $Mg_2SiO_4$ retains its solid state at the same temperature, with a thermal conductivity of approximately 14 W/mK, which is still more than four times that of $Mg_2SiO_4He_2$. These results indicate that the incorporation of He significantly reduces the thermal conductivity of $Mg_2SiO_4$ at high temperatures, particularly after the superionic transition. This pronounced reduction in thermal conductivity strongly supports the formation of a thermal boundary layer at the hot-Jupiter core-mantle boundary.

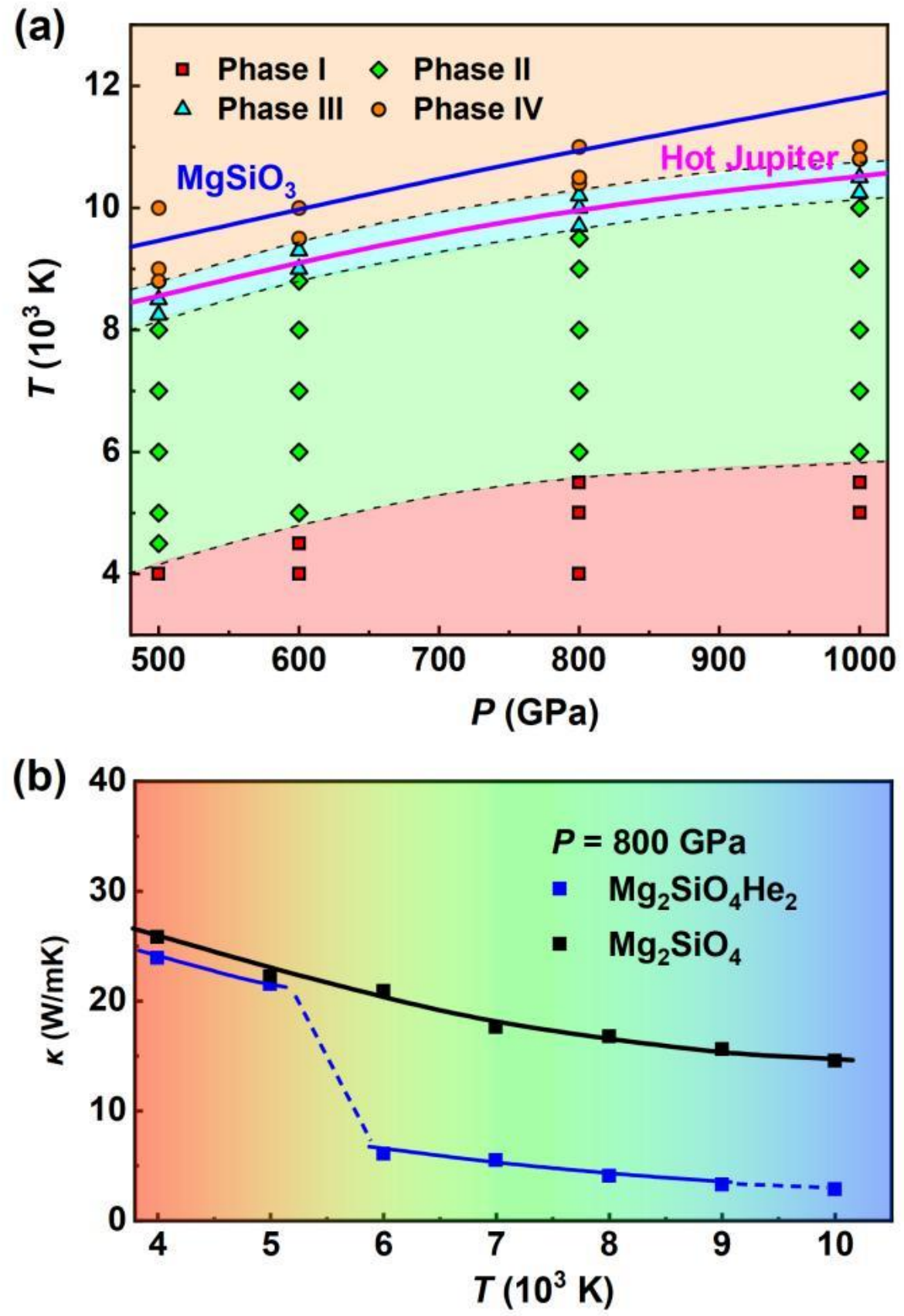

**Fig.3** **a** *P*-*T* phase diagram of $Mg_2SiO_4He_2$ from MLMD simulations. The black dashed line is the fitted phase boundary; the blue and pink solid lines are the melting line of $MgSiO_3$ [44] and the *P*-*T* curves of hot Jupiters[10, 41, 42], respectively. **b** Temperature-dependent thermal conductivity of $Mg_2SiO_4He_2$ and $Mg_2SiO_4$ at 800 GPa. The background colors represent the transition of $Mg_2SiO_4He_2$ from the solid (red) to the superionic state phase II (green) and phase III (blue), and $Mg_2SiO_4$ always exists in the solid state.

## Discussion

Based on these findings, we propose that $Mg_2SiO_4He_2$ formation provides a chemical mechanism for delayed cooling of hot Jupiters. At core-mantle boundary of a hot Jupiter (pressure > 700 GPa), helium (accreted with the primordial nebular gas or delivered by helium rain) in the planetary rock core and deep mantle spontaneously reacts with $Mg_2SiO_4$ to form $Mg_2SiO_4He_2$ compounds, as shown in Fig. 4. The temperature at this boundary is sufficiently high to drive He atoms into the superionic and liquid states, maximizing their diffusion rate and enabling rapid helium diffusion through the lattice channels. Crucially, the chemical incorporation of helium from the surrounding metallic hydrogen envelope into the silicate phase removes helium from the convecting gas, creating a compositional gradient (helium-depleted above, helium-enriched below) across the core-mantle boundary. Such compositional stratification can inhibit large-scale convection[45] and establish a thermal boundary layer across which heat transport is governed by conduction rather than convection, substantially reducing the cooling flux from the deep interior. In a planet where the core-mantle boundary region is chemically stratified by $Mg_2SiO_4He_2$ formation, convective heat transport from the deep interior to the outer envelope is impeded. The core therefore retains its primordial heat for longer, slowing the planet's contraction. This "delayed cooling" effect keeps the planet at a radius larger than standard thermal evolution models predict. Observational data confirm that the radius anomaly of hot Jupiters is closely linked to their cooling history: slower cooling corresponds to larger radii[46, 47]. The $Mg_2SiO_4He_2$ compositional stratification mechanism provides a concrete chemical basis for this delayed cooling, complementing the shallow heating mechanisms (such as Ohmic dissipation) that drive reinflation of older planets.

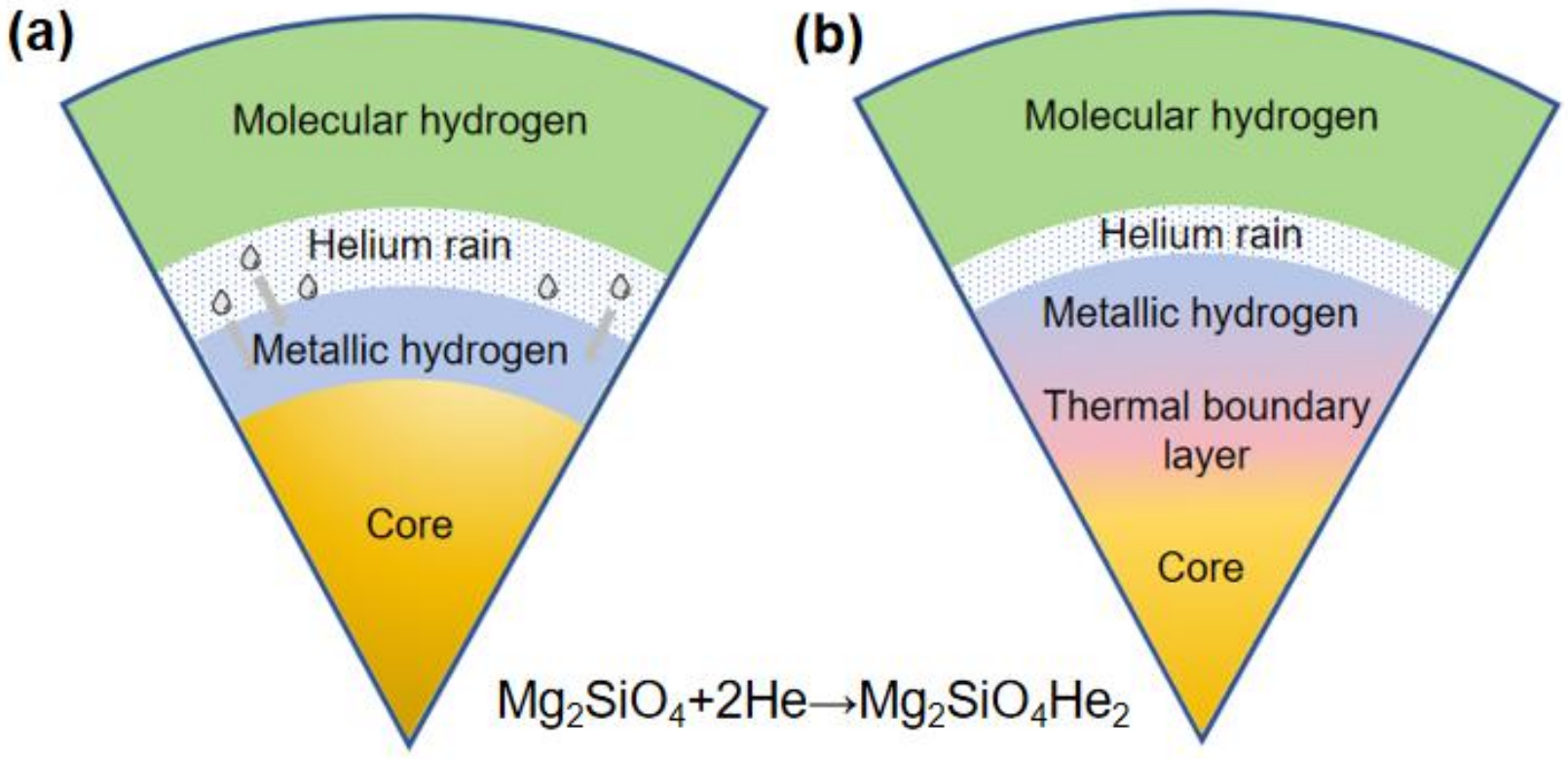

**Fig.4 a** The classical multi-layer model of a hot Jupiter. **b** A revised model incorporating silicate and helium reactions, including a novel thermal boundary layer that impedes the loss of internal heat.

Current models of hot Jupiter radius inflation broadly divide into active heating mechanisms that can reinflate cooled planets[41, 42], and delayed-cooling mechanisms that slow the initial contraction but cannot reinflate. Recent population-level analyses suggest that both components operate in concert[47]. Our proposed mechanism falls naturally into the delayed-cooling category: the formation of $Mg_2SiO_4He_2$ at the core-mantle boundary creates a chemically stratified layer that inhibits convective heat loss from the deep interior. This compositional barrier complements atmospheric and shallow heating processes. Enhanced atmospheric opacity creates a thermal boundary layer near the photosphere that slows radiative cooling; the $Mg_2SiO_4He_2$ mechanism establishes a thermal boundary layer at the base of the envelope, where its effects on the overall thermal evolution are potentially more significant because it directly impedes convective heat loss from the core. The two mechanisms operate at opposite ends of the planet's radial structure and are not mutually exclusive. The efficiency of this mechanism depends on the pressure-temperature conditions and on the availability of helium at depth, whether delivered by helium rain in cooler hot Jupiters where H-He immiscibility operates, or present from primordial mixing during formation as suggested by Juno's observation of Jupiter's dilute core[48-50]. These dependencies provide concrete targets for future interior modeling and observational validation.

# Conclusion

In summary, our study predicts a novel helium-silicate compound $Mg_2SiO_4He_2$ that is thermodynamically stable from ~577 GPa up to at least 1000 GPa and elucidates its stepwise melting behavior. Helium incorporation yields a compound whose superionic transition lies far below the melting point of the helium-free silicate $Mg_2SiO_4$. The chemical sequestration of helium into this compound at the core-mantle boundary creates compositional stratification that can inhibit convective heat transport and effectively delay core cooling, providing a chemical mechanism for delayed cooling that complements existing models of radius inflation of hot Jupiters. This work predicts the first stoichiometrically stable Mg-bearing silicate-helium compound, providing critical insights into silicate-noble-gas interactions and the thermal evolution of exoplanetary interiors.

# Methods

The AIRSS code[51] was employed for structure searches in the $Mg_2SiO_4$-He system at different pressures and the CASTEP code[52] was used for structural relaxation with a lower calculated accuracy. The cutoff energy and k-point grid were set to 800 eV and $2\pi\times0.03$ Å$^{-1}$ for high-precision structural relaxation calculations.

The electronic structures were calculated in the VASP code[53], with a truncation energy set at 1000 eV. The exchange-correlation functional was the Perdew-Burke-Ernzerhof (PBE) method within the generalized gradient approximation (GGA) framework[54]. The k-point meshes with grid spacing of $2\pi \times 0.02$ Å$^{-1}$ were used for high-accuracy structural relaxations and electronic properties calculations. The PHONOPY code[55, 56] was utilized to calculate phonon spectrum.

To investigate the thermodynamic behavior and phase transitions under high temperatures, a machine-learning interatomic potential was constructed using the DeePMD-kit code[57-59]. For capturing the different phases in the temperature-pressure phase diagram, many dynamic trajectories were accumulated from the Ab initio molecular dynamics simulations by using the VASP code. Molecular dynamics simulations were performed using the LAMMPS package[60, 61] under the NPT ensemble. Temperature was controlled using a Nosé-Hoover thermostat[62]. Solid, superionic, and liquid phases were distinguished by analyzing the MSD of individual atoms. The thermal conductivity was calculated by using the Green-Kubo formula[63, 64]. The detailed calculation formula is provided in SI.

# References

1. Mayor, M., & Queloz, D. A Jupiter-mass companion to a solar-type star, *Nature* **378,** 355 (1995).

2. Charbonneau, D., Brown, T.M., Latham, D.W., & Mayor, M. Detection of Planetary Transits Across A Sun-Like Star, *Astrophys. J.* **529,** L45 (2000).

3. Henry, G.W., Marcy, G.W., Butler, R.P., & Vogt, S.S. A Transiting "51 Peg–Like" Planet, *Astrophys. J.* **529,** L41 (2000).

4. Miller, N., & Fortney, J.J. The Heavy-Element Masses of Extrasolar Giant Planets, Revealed, *Astrophys. J.* Lett. **736,** L29 (2011).

5. Demory, B.-O., & Seager, S. Lack of Inflated Radii for Kepler Giant Planet Candidates Receiving Modest Stellar Irradiation, *Astrophys. J. Suppl. S.* **197,** 12 (2011).

6. Weiss, L.M., et al. The Mass of KOI-94d and a Relation for Planet Radius, Mass, and Incident Flux, *Astrophys. J.* **768,** 14 (2013).

7. Bodenheimer, P., Lin, D.N.C., & Mardling, R.A., On the Tidal Inflation of Short-Period Extrasolar Planets, *Astrophys. J.* **548,** 466 (2001).

8. Batygin, K., & Stevenson, D.J. Inflating Hot Jupiters with Ohmic Dissipation, *Astrophys. J. Lett.* **714,** L238 (2010).

9. Burrows, A., Hubeny, I., Budaj, J., & Hubbard, W.B. Possible Solutions to the Radius Anomalies of Transiting Giant Planets, *Astrophys. J.* **661,** 502 (2007).

10. Wu, Y., & Lithwick, Y. Ohmic Heating Suspends, Not Reverses, the Cooling Contraction of Hot Jupiters, *Astrophys. J.* **763,** 13 (2013).

11. Thorngren, D.P., & Fortney, J.J. Bayesian Analysis of Hot-Jupiter Radius Anomalies: Evidence for Ohmic Dissipation? *Astrophys. J.* **155,** 214 (2018).

12. Sarkis, P., Mordasini, C., Henning, T., Marleau, G.D., & Mollière, P. Evidence of three mechanisms explaining the radius anomaly of hot Jupiters, *Astron. Astrophys.* **645,** A79 (2021).

13. Tschauner, O., Ma, C., Beckett, J.R., Prescher, C., Prakapenka, V.B., & Rossman, G.R. Discovery of bridgmanite, the most abundant mineral in Earth, in a shocked meteorite, *Science* **346,** 1100 (2014).

14. Wu, S.Q., et al. An adaptive genetic algorithm for crystal structure prediction, *J. Phys.: Condens. Matter* **26,** 035402 (2014).

15. Umemoto, K., Wentzcovitch, R.M., Wu, S., Ji, M., Wang, C.-Z., & Ho, K.-M. Phase transitions in $MgSiO_3$ post-perovskite in super-Earth mantles, *Earth Planet. Sci. Lett.* **478,** 40 (2017).

16. Ritterbex, S., Harada, T., & Tsuchiya, T. Vacancies in MgO at ultrahigh pressure: About mantle rheology of super-Earths, *Icar* **305,** 350 (2018).

17. Deng, J., & Stixrude, L. Thermal Conductivity of Silicate Liquid Determined by Machine Learning Potentials, *Geophys. Res. Lett.* **48,** 17 (2021).

18. Tschauner, O., et al. Discovery of davemaoite, $CaSiO_3$-perovskite, as a mineral from the lower mantle, *Science* **374,** 891 (2021).

19. Pan, S., et al. Magnesium oxide-water compounds at megabar pressure and implications on planetary interiors, *Nat. Commun.* **14,** 1165 (2023).

20. Yin, K., Belonoshko, A.B., Li, Y., & Lu, X. Davemaoite as the mantle mineral with the highest melting temperature, *Sci. Adv.* **9,** eadj2660 (2023).

21. Wang, Z., He, Y., Mao, H.-K., & Kim, D.Y. Superionicity of oxygen-deficient davemaoite and its impact on the deep-Earth oxidation cycle, *Sci. Adv.* **11,** eadu8401 (2025).

22. Feng, X., et al. Nanosecond structural evolution in shocked coesite, *Sci. Adv.* **11,** eads3139 (2025).

23. Zhang, Y., et al. Collective motion in hcp-Fe at Earth's inner core conditions, *Proc. Natl. Acad. Sci. USA* **120,** e2309952120 (2023).

24. Zhang, Y., Wanga, W., Li, Y., & Wu, Z. Superionic iron hydride shapes ultralow-velocity zones at Earth's core-mantle boundary, *Proc. Natl. Acad. Sci. USA* **121,** e2406386121 (2024).

25. Wang, X., et al. Pressure-induced redox reversal of iron and the distribution of elements in deep Earth, *Proc. Natl. Acad. Sci. USA* **122,** e2414911122 (2025).

26. Wu, Z., et al. Ab Initio Superionic-Liquid Phase Diagram of $Fe_{1-x}O_x$ under Earth's Inner Core Conditions, *Phys. Rev. Lett.* **136,** 016103 (2026).

27. Deng, J., Gou H., & Hu, Q. Hydrogen-carbon doubly superionic conduits of carbonic acids in planetary ices, *Sci. Adv.* **12,** eaeb9271 (2026).

28. Gao, H., et al. Coexistence of plastic and partially diffusive phases in a helium-methane compound, *Natl. Sci. Rev.* **7,** 1540 (2020).

29. Liu, C., et al. Plastic and Superionic Helium Ammonia Compounds under High Pressure and High Temperature, *Phys. Rev. X* **10,** 021007 (2020).

30. Liu, C., et al. Partially Diffusive Helium-Silica Compound under High Pressure, *Chinese Phys. Lett.* **39,** 076101 (2022).

31. Zhang, J., Liu, H., Chen, C., & Ma, Y. Sequestration of helium and xenon via iron-halide compounds in early Earth, *Matter Radiat. Extrem.* **9,** 037403 (2024).

32. Bai, Y., et al. Electrostatic force driven helium insertion into ammonia and water crystals under pressure, *Commun. Chem.* **2,** 102 (2019).

33. Liu, Z., et al. Reactivity of He with ionic compounds under high pressure, *Nat. Commun.* **9,** 951 (2018).

34. Dong, X., et al. A stable compound of helium and sodium at high pressure, *Nat. Chem.* **9,** 440 (2017).

35. Takezawa, H., et al. Formation of Iron-Helium Compounds under High Pressure, *Phys. Rev. Lett.* **134,** 084101 (2025).

36. Liu, C., et al. Multiple superionic states in helium-water compounds, *Nat. Phys.* **15,** 1065 (2019).

37. Gao, H., et al. Superionic Silica-Water and Silica-Hydrogen Compounds in the Deep Interiors of Uranus and Neptune, *Phys. Rev. Lett.* **128,** 035702 (2022).

38. Guo, L., & Zhang, J. New MgO-$H_2O$ compounds at extreme conditions, *Chinese Phys. B* **32,** 076201 (2023).

39. Shao, S., et al. New $MgSiO_4H_2$ Phases as Potential Primary Water Carriers into the Deep Earth, *Phys. Rev. Lett.* **133,** 214101 (2024).

40. Li, H.-F., Oganov, A.R., Cui, H., Zhou, X.-F., Dong, X., & Wang, H.-T. Ultrahigh-Pressure Magnesium Hydrosilicates as Reservoirs of Water in Early Earth, *Phys. Rev. Lett.* **128,** 035703 (2022).

41. Komacek, T.D., & Youdin, A.N. Structure and Evolution of Internally Heated Hot Jupiters, *Astrophys. J.* **844,** 94 (2017).

42. Komacek, T.D., Thorngren, D.P., Lopez, E.D., & Ginzburg, S. Reinflation of Warm and Hot Jupiters, *Astrophys. J.* **893,** 36 (2020).

43. Militzer, B., & Hubbard, W.B. Study of Jupiter's interior: Comparison of 2, 3, 4, 5, and 6 layer models, *Icar* **411,** 115955 (2024).

44. Fei, Y., et al. Melting and density of $MgSiO_3$ determined by shock compression of bridgmanite to 1254 GPa, *Nat. Commun.* **12,** 876 (2021).

45. Chabrier, G., & Baraffe, I. Heat Transport in Giant (EXO) Planets: a New Perspective, *Astrophys. J.* **661,** L81 (2007).

46. Thorngren, D.P., Fortney, J.J., Lopez, E.D., Berger, T.A., & Huber, D. Slow Cooling and Fast Reinflation for Hot Jupiters, *Astrophys. J. Lett.* **909,** L16 (2021).

47. Schmidt, S.P., Thorngren, D.P., & Schlaufman, K.C. Hot Jupiters Are Inflated Primarily by Shallow Heating, *Astrophys. J.* **1001,** 35 (2026).

48. Wahl, S.M., et al. Comparing Jupiter interior structure models to Juno gravity measurements and the role of a dilute core, *Geophys. Res. Lett.* **44,** 4649 (2017).

49. Brygoo, S., et al. Evidence of hydrogen-helium immiscibility at Jupiter-interior conditions, *Nature* **593,** 517 (2021).

50. Wilson, H.F., & Militzer, B. Sequestration of Noble Gases in Giant Planet Interiors, *Phys. Rev. Lett.* **104,** (2010).

51. Pickard, C.J., & Needs, R.J. Ab initio random structure searching, *J. Phys.: Condens. Matter* **23,** 053201 (2011).

52. Segall, D. M., et al. First-principles simulation: ideas, illustrations and the CASTEP code, *J. Phys.: Condens. Matter* **14,** 2717 (2022).

53. Kresse, G., & Furthmüller, J. Efficient iterative schemes for ab initio total-energy calculations using a plane-wave basis set, *Phys. Rev. B* **54,** 16 (1996).

54. John, P.P., Kieron, B., & Matthias, E. Generalized Gradient Approximation Made Simple, *Phys. Rev. Lett.* **77,** 18 (1996).

55. Togo, A., Chaput, L., Tadano, T., & Tanaka, I. Implementation strategies in phonopy and phono3py, *J. Phys.: Condens. Matter* **35,** 353001 (2023).

56. Togo, A. First-principles Phonon Calculations with Phonopy and Phono3py, *J. Phys. Soc. Jpn.* **92,** 012001 (2023).

57. Wang, H., Zhang, L., Han, J., & E, W. DeePMD-kit: A deep learning package for many-body potential energy representation and molecular dynamics, *Comput. Phys. Commun.* **228,** 178 (2018).

58. Zeng, J., et al. DeePMD-kit v2: A software package for deep potential models, *J. Chem. Phys.* **159,** 054801 (2023).

59. Zeng, J., et al. DeePMD-kit v3: A Multiple-Backend Framework for Machine Learning Potentials, *J. Chem. Theory Comput.* **21,** 4375 (2025).

60. Plimpton, S. Fast Parallel Algorithms for Short-Range Molecular Dynamics, *J. Comput. Phys.* **117,** 1 (1995).

61. Thompson, A.P., et al. LAMMPS - a flexible simulation tool for particle-based materials modeling at the atomic, meso, and continuum scales, *Comput. Phys. Commun.* **271,** 108171 (2022).

62. Nosé, S. A unified formulation of the constant temperature molecular dynamics methods, *J. Chem. Phys.* **81,** 511 (1984).

63. Surblys, D., Matsubara, H., Kikugawa, G., & Ohara, T. Methodology and meaning of computing heat flux via atomic stress in systems with constraint dynamics, *J. Appl. Phys.* **130,** 215104 (2021).

64. Surblys, D., Matsubara, H., Kikugawa, G., & Ohara, T. Application of atomic stress to compute heat flux via molecular dynamics for systems with many-body interactions, *Phys. Rev. E* **99,** 051301 (2019).